\documentclass{article}
\usepackage{mrs2005,epsfig}
\begin{document} 
\title{FAINT FAR-INFRARED SOURCES: GALAXIES, CLUSTERS, OR CLUSTER GALAXIES}

\author{Petri V\"ais\"anen}
\affil{South African Astronomical Observatory, Cape Town, South Africa} 

\author{Mika Juvela, Kalevi Mattila}
\affil{Observatory, University of Helsinki, Helsinki, Finland}

\author{Jari Kotilainen}
\affil{Tuorla Observatory, University of Turku, Piikki\"o, Finland}

\begin{abstract} 
We present results from an on-going follow-up campaign of far-infrared 
sources detected as part of our ISOPHOT Cosmic IR Background project.
Fields have been imaged in the optical and near-infrared, and we find
at least a third of the FIR targets areas to contain a bright and nearby 
star-forming galaxy. 
We also explore the largely neglected possibility that instead of individual
galaxies some of the fainter FIR sources are confused sums of several
sources -- or even whole cores of galaxy clusters at redshifts of
$z\sim0.4-0.8$.  We look for correlations in the FIR positions with
extremely red objects (EROs) and significant peaks in the galaxy surface
density and peaks in cluster red sequence signal.  Several matches are found
and we have set out to study cluster candidates spectroscopically.  
The campaign is producing an interesting base to study IR-luminous, 
strongly star-forming galaxies in potential cluster environments.
\end{abstract} 
 
\section{Introduction} 
 
The Cosmic IR background (CIRB) consists of the integrated light of all 
galaxies along the line-of-sight plus any intergalactic contributions, 
and includes up to 2/3 of all energy generated since recombination 
(\cite{Puget96, Elbaz03}). 
Starlight lost to dust obscuration reappears in the far-IR CIRB after 
reprocessing in galaxies, especially during their dusty formation era, 
making the CIRB an important window into the enigmatic era of galaxy formation.

Much effort has been put into resolving the CIRB and studying the galaxies 
responsible for it (e.g. \cite{Dole01}). 
Because of the large beam ($\approx1$ arcmin) of the 
ISO and Spitzer at FIR, the unambiguous ground-based identification and 
follow-up of such faint FIR galaxies has been extremely difficult.
In the ISOPHOT EBL project (\cite{Juvela00}) we detected 55 FIR 
objects at multiple wavelengths from 90 to 180 $\mu$m down to $\sim$100 mJy.  
All detections are in at least two FIR bands allowing the separation of 
galactic cirrus knots and helping to fit galaxy templates of varying 
temperatures and redshifts.

Optical and near-infrared follow-up of the sources has been recently performed 
with the Nordic Optical Telescope and the VLT and the areas are all covered 
with R, I, and K imaging to R$\approx$24 and K$\approx$20.  We have also 
started multi-object spectroscopic follow-up of selected fields.  
Here we present results from the North Galactic Pole hole ($l\approx88$, 
$b\approx73$ that has practically as low hydrogen column density as the 
Lockman hole.

\section{Star-forming galaxies and confused sources}

Of the 22 individual ISOPHOT detections in the NGP field, we observed 21. 
Of these 21 targets, 7 are found to be unambiguous, relatively nearby disk 
galaxies.  We fit a range of SED templates to the photometry from the optical 
to FIR and find these galaxies to be well fit by evolved Sc or NGC 6090 
starburst templates at redshifts of z = 0.05 -- 0.25 (see left panel of Fig.1).

%
\begin{figure}  
\vspace*{0.01cm}  
\begin{center}
\epsfig{figure=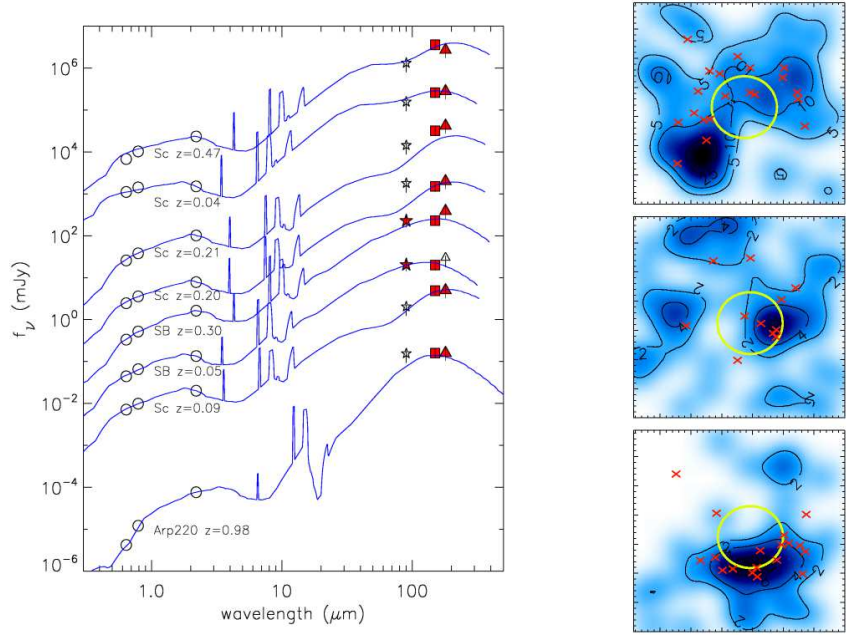,width=13.8cm}  
\end{center}
\vspace*{0.01cm}  
\caption{{\em Left panel}: Observed and fitted SEDs of 7 bright disk galaxies
(shifted for clarity); 
redshifts and types are indicated.  The lowest SED is an example of our 
fainter red sources, with a ULIRG template, found in many target fields.
{\em Right panel}:
Large circles are FIR targets. The map and contours show the 
surface overdensities of galaxies with respect to field galaxies, 
in colour bins of $R-I=1.3-1.5$, $1.1-1.3$, and $1.5-1.7$, from top to bottom.
Small red crosses are EROs (R-K$>5$). 
} 
\end{figure} 

We can not find unambiguous identifications for the remaining 14 NGP FIR 
sources, even though 7 of them do have relatively bright (R$\approx$20 mag) 
galaxies in the field.  In a typical case there are two or more such bright 
galaxies, but none of them alone can easily explain the FIR source, while a 
combination could. In addition, a total of 8 of these fields have one or more 
faint red sources whose optical-to-FIR SEDs are well fit with an 
ultra-luminous IR-galaxy SED template, such as Arp 220, at z$\sim$1.  

If the FIR flux of half of these confused sources are split into two separate 
FIR sources, the FIR source counts in the range 100-200 mJy reduce by 50\%.  
If the remaining sources were split into 3 separate sources, source 
counts in this range decrease by a factor of 2.  Since model fits assume 
single galaxies to populate FIR source count bins, this confusion has 
potentially serious effects on the interpretation of FIR galaxy counts.   

\section{Clusters of galaxies}

Instead of attempting to force individual optical and NIR counterparts for 
each FIR source, we also searched for correlations in their positions with 
extremely red objects (EROs) and significant peaks in the galaxy surface 
density, and peaks in cluster red sequence signal.  The motivation is that 
some faint ISOPHOT sources could in principle be caused by the integrated 
emission of whole galaxy clusters.  Our 
simulations show that combined radiation of normal galaxies in the central 
areas of rich clusters at redshifts $z=0.4-0.8$ result in point-sources of 
50-200 mJy in the FIR when seen through the ISOPHOT PSF.

Indeed, 4 of the 14 NGP FIR fields without an unambiguous bright galaxy 
counterpart can be associated with $> 3 \sigma$ ERO overdensities. Moreover, 
6 of the 14 fields are associated with overdensities of $> 6 \sigma$ in galaxy 
surface density calculated in R-I colour slices (right panel of Fig.1).  
One of the identified cluster candidates is a previously detected 
X-ray cluster at $z=0.70$. We are currently studying the FIR fields 
spectroscopically for cluster confirmations.

\section{Conclusions} 
 
Classification of 21 NGP hole FIR sources from ISOPHOT CIRB project yields
7 unambiguous nearby star-forming galaxies and/or starbursts.  The rest are
confused targets, with likely contribution from 2-3 bright nearby star-forming 
galaxies in half the cases and possible contribution from z$\sim 1$ ULIRGs 
in 5-7 cases. The confusion of FIR sources has significant consequences on 
the interpretation of FIR source counts. 

We also find significant spatial correlation of FIR positions of the confused 
target class with ERO overdensities and red colour sequences.



\vfill 

\begin{thebibliography}{}{ 
\begin{tabular}{lll}
\begin{minipage}[l]{0.45\textwidth}
\bibitem{Dole01}  Dole et al., 2001, A\&A 372, 364
\bibitem{Elbaz03}  Elbaz \& Cesarsky, 2003, Science 300, 270
\end{minipage}
&
&
\begin{minipage}[r]{0.45\textwidth}
\bibitem{Juvela00}  Juvela et al., 2000, A\&A 360, 813
\bibitem{Puget96}  Puget et al., 1996, A\&A 308, L5
\end{minipage}
\end{tabular}
} 
\end{thebibliography}
\end{document}